\documentclass[a4paper,conference]{IEEEtran}

\usepackage{cite}
\usepackage{url}

\usepackage{amsmath,amssymb,amsfonts}
\usepackage{algorithmic}
\usepackage{graphicx}
\usepackage{xcolor}
\usepackage{bm}

\usepackage{optidef}
\usepackage{amsthm}
\usepackage{caption}
\usepackage{subcaption}
\newcommand{\E}{\mathbb{E}}
\newcommand{\cS}{\mathcal{S}}
\newcommand{\cA}{\mathcal{A}}
\newcommand{\cB}{\mathcal{B}}

\newtheorem{lemma}{Lemma}

\newcommand{\realizationY}{\hat{y}}

\usepackage{comment}
\usepackage{lipsum}
\usepackage{algorithm}
\usepackage{svg}
\usepackage{caption}

\usepackage{listings}
\usepackage{xcolor}

\definecolor{ReviewerColorA}{rgb}{0.40, 0.30, 0.90}
\definecolor{ReviewerColorB}{rgb}{1.00, 0.55, 0.00}
\definecolor{ReviewerColorC}{rgb}{0.00, 0.45, 0.90}
\definecolor{ReviewerColorD}{rgb}{0.00, 0.65, 0.30}
\definecolor{brightmaroon}{rgb}{0.76, 0.13, 0.28}

\newcommand{\ReviewerA}[1]{\textcolor{ReviewerColorA}{Reviewer 1: #1}}
\newcommand{\ReviewerB}[1]{\textcolor{ReviewerColorB}{Reviewer 2: #1}}
\newcommand{\ReviewerC}[1]{\textcolor{ReviewerColorC}{Reviewer 3: #1}}
\newcommand{\ReviewerD}[1]{\textcolor{ReviewerColorD}{Reviewer 4: #1}}
\newcommand{\alex}[1]{{\color{red} ALEX: #1}}
\newcommand{\esr}[1]{{\color{brightmaroon} ESR: #1}}

\usepackage{ifthen}

\renewcommand{\ReviewerA}[1]{}
\renewcommand{\ReviewerB}[1]{}
\renewcommand{\ReviewerC}[1]{}
\renewcommand{\ReviewerD}[1]{}
\renewcommand{\alex}[1]{}
\renewcommand{\esr}[1]{}

\definecolor{codegreen}{rgb}{0,0.6,0}
\definecolor{codegray}{rgb}{0.5,0.5,0.5}
\definecolor{codepurple}{rgb}{0.58,0,0.82}
\definecolor{backcolour}{rgb}{0.95,0.95,0.92}

\lstdefinestyle{mystyle}{
backgroundcolor=\color{backcolour}, 
    commentstyle=\color{codegreen},
    keywordstyle=\color{magenta},   numberstyle=\tiny\color{codegray},
    stringstyle=\color{codepurple},
basicstyle=\ttfamily\footnotesize,
    breakatwhitespace=false,        
    breaklines=true,                
    captionpos=b,                   
    keepspaces=true,                
    numbers=left,                   
    numbersep=5pt,                  
    showspaces=false,               
    showstringspaces=false,
    showtabs=false,              
    tabsize=2
}

\newcommand{\mysup}[0]{^{(N)}}

\begin{document}


\newboolean{compactcameraready}
\setboolean{compactcameraready}{false} 

\newcommand{\notes}[1]{\ifincludeadditionalnotes #1\fi}


\ifthenelse{\boolean{compactcameraready}}{
}{
}


\title{Online Learning of Weakly Coupled MDP Policies for Load Balancing and Auto Scaling}




\author{\IEEEauthorblockN{S.R. Eshwar\IEEEauthorrefmark{1}, Lucas Lopes Felipe\IEEEauthorrefmark{4}, Alexandre Reiffers-Masson\IEEEauthorrefmark{3}, Daniel Sadoc {\color{black}Menasché}\IEEEauthorrefmark{4}, Gugan Thoppe\IEEEauthorrefmark{1}}
\IEEEauthorblockA{\IEEEauthorrefmark{1}Department of Computer Science and Automation, Indian Institute of Science, Bangalore \\ \IEEEauthorrefmark{3}Department of Computer Science, 
IMT Atlantique Bretagne-Pays de la Loire, 
Campus de Brest, 
29238 Brest CEDEX 03 \\ \IEEEauthorrefmark{4}Institute of Computing, Federal University of Rio de Janeiro (UFRJ), Rio de Janeiro, Brazil }}

\IEEEoverridecommandlockouts
\IEEEpubid{\makebox[\columnwidth]{ISBN 978-3-903176-63-8 \copyright 2024 IFIP\hfill}
\hspace{\columnsep}\makebox[\columnwidth]{ }}

\maketitle

\IEEEpubidadjcol 






\begin{abstract}
Load balancing and auto scaling are at the core of scalable,   contemporary systems, addressing dynamic resource allocation and service rate adjustments in response to workload changes.  This paper introduces a novel model and algorithms for   tuning   load balancers coupled with auto scalers, considering bursty traffic arriving at finite queues. We begin by presenting the problem as a weakly coupled Markov Decision Processes (MDP), solvable via a  linear program (LP).  However, as the number of  control variables of such LP grows combinatorially,  we introduce a more tractable relaxed  LP formulation, and extend it to  tackle the problem of  online parameter learning and policy optimization using a two-timescale algorithm based on the LP Lagrangian. 
%
Our numerical experiments shed insight into properties of the optimal policy. In particular, we identify a phase transition in  the probability of job acceptance as a function of the job dropping costs. 
 The experiments also indicate the efficacy of the proposed  online learning method, that learns parameters together with the optimal policy, in converging to the optimal solution of the relaxed LP. In summary, the contributions of this work encompass an analytical model and its LP-based solution approach, together with an online learning algorithm,  offering insights into the effective management of  distributed systems.
\end{abstract}


\section[Introduction]{Introduction}
\label{sec:introduction}

The management of distributed systems has grown increasingly complex with the widespread adoption of cloud computing and microservice architectures. These architectures rely on two fundamental pillars: load balancing, which dynamically distributes traffic among servers, and auto scaling, 
which adjusts service rates in response to workload changes. 
Load balancers and auto scalers are pervasive in numerous modern systems, ranging from routers to databases and web systems~\cite{imp_lb_as_in_cloud,aws_auto_scaling,aws_elb}. 

Despite the extensive literature on load balancing and auto scaling methods, most studies tend to treat these elements separately~\cite{ml_lb_survey1, ml_lb_survey2, ml_lb_survey3, rl_as_survey} or focus on formal results for asymptotic regimes~\cite{chapter4}. However, 
 as systems evolve over time due to workload changes, such as in e-commerce platforms experiencing surges in user traffic during peak seasons or promotional events, the need for rapid resource provisioning and efficient utilization to prevent service disruptions becomes paramount. This paper addresses the concurrent optimization of load balancing and auto scaling in such scenarios.

In this paper, we introduce a novel model and algorithm addressing the general problem of load balancing and auto scaling in a system of finite parallel queues with unknown parameters and bursty traffic. We present the problem as a finite horizon weakly coupled Markov Decision Process (MDP), where decisions on load balancing and auto scaling are taken at each arrival of a bulk of jobs. These decisions involve determining which queues will receive new jobs and at what rate each queue will serve them. The weakly coupled MDP solution is achieved through a linear program (LP) (see Section \ref{subsec:lp-formulation}). However, it faces challenges due to the combinatorial growth of optimization variables. This motivates a relaxation of the problem by replacing instantaneous constraints by constraints on expectations. As a result, a simpler LP formulation is obtained (see Section \ref{subsec:relaxed-lp}).

Solving the proposed LP requires knowledge of system dynamics, which may not always be available upfront. To address this, we propose a sample-based online learning algorithm to find the optimal policy for load balancing and auto scaling. This approach involves using a two-timescale algorithm to solve the LP by considering its Lagrangian (see Section \ref{subsec:2ts-sa}). Numerical experiments (see Section \ref{sec:experiments}) across various scenarios demonstrate the convergence of the proposed method to the optimal LP solution, indicating its effectiveness in addressing the challenges of load balancing and auto scaling in dynamic systems with bursty traffic.

\textbf{Pior art. } 
Given the widespread adoption of load balancing and auto scaling, a few   frameworks considered a stochastic model for their joint analysis (see~\cite{ALBA,optimal_service_elasticity,JIQ_service_elasticity,monitorless_LBA}).  Our work also considers load balancing and auto scaling decisions. However, by tackling the problem as a weakly coupled MDP, we can embrace general dispatching and auto scaling rules, beyond the ones considered in the previous works.  In this work, we leverage~\cite{lp-policy,gast2021lp} showing its applicability in optimizing resource allocation and service rates in distributed systems. {\color{black}In addition, we extend those previous works by proposing a method to learn LP-based policies, accounting for the case when parameters are unknown.}  

\textbf{Contributions. }
In summary, our main contributions are:


  
  \textbf{Analytical model:} We formulate the load balancing and service rate control problem as a Weakly Coupled MDP (WC-MDP). The objective is to minimize costs arising from delays (proportional to queue lengths), energy consumption (proportional to service rate), and job dropouts due to full queues. These factors are integrated into an LP along with workload and resource constraints (Sections~\ref{sec:model},~\ref{subsec:opt-problem}, and~\ref{subsec:lp-formulation}). Note that, by treating the joint load balancing and auto scaling problem as a weakly coupled MDP, we can adopt more general dispatching and auto scaling rules beyond traditional strategies.

   
   \textbf{Online learning algorithm:} We utilize recent advancements in LP-based policies to address a relaxed version of the above mentioned LP problem (Section~\ref{subsec:relaxed-lp}). Recognizing that system parameters may not be initially observable, we devise an online learning algorithm to approximate the LP-based policy. Specifically, we leverage the Lagrangian of the LP in this process. Our algorithm employs a two-timescale stochastic approximation, where Lagrange multipliers are computed based on the current estimate of control variables, and then control variables are recalculated given the Lagrange multipliers (Section~\ref{subsec:2ts-sa}). Notably, this approach enables decision-making at each iteration of the algorithm, aligning with an online operational paradigm.

  
  \textbf{Numerical experiments:} We illustrate   properties of optimal policies, such as a phase transition in job acceptance probability, indicating a shift from predominant   rejection to full acceptance as rejection costs increase (Section~\ref{sec:experiments}).  We also  discuss convergence properties of the proposed algorithms.  

\textbf{Outline.} The remainder of this paper is organized as follows. In the upcoming section, we present   background and related work. Then, Section~\ref{sec:model} introduces the proposed model,  Section~\ref{sec:problem-formulation}  contains our problem formulation and solution, Section~\ref{sec:experiments} reports our results and  Section~\ref{sec:conclusion} concludes.

\section[Model]{Model} \label{sec:model}

Next, 
we detail the considered queuing system  and its corresponding  dynamics.  We also introduce basic notation.

\label{subsec:queuing-system}

The \textbf{system} is composed of $N$ queues designated for the processing of jobs, each equipped with a finite buffer of capacity $K$.  
The \textbf{controller} has two controls for each of the $N$ queues: (1) \emph{load balancing} involves deciding whether to send a job to the tail of the queue; (2) \emph{service rate management} 
determines the service rate of the queue, which in turn impacts the  likelihood of  processing a job in a given time slot, i.e., the probability of a service completion at a given  slot is a function of whether the high or low service rates were chosen.

\ReviewerB{"which in turn impacts the likelihood of successful job processing" is misleading.  All processing is successful, the question is whether a processing is completed in that timeslot, or not.} \esr{- Done}

The \textbf{job arrival process} consists of  arrivals of batches of $\alpha N$  jobs, where $0 < \alpha < 1$. 
The arrival time of the $i$-th batch of new jobs is denoted as $T_i\in\mathbb{N}_+$. The inter-arrival time, $T_i - T_{i-1}$, follows a geometric distribution with parameter $p$, given by $\mathbb{P}(T_i-T_{i-1}=\tau)=(1-p)^{\tau-1}p,\;\forall \tau\geq~1$.


The \textbf{dynamics of each queue} is characterized at embedded points, corresponding to batch arrivals. 
Let $S_n(T_{i})$ denote the number of jobs in the $n$-th queue   at the beginning of slot $T_i$. We denote by  $A_n(T_{i})$ an indicator variable, equal to 1 if a job is admitted to the $n$-th queue at  $T_{i}$ and 0 otherwise. Let   $D_{n}(T_{i})$ be the number of jobs processed between $T_{i}$ and $T_{i+1}$. 

\ReviewerB{it seems that one job can be forwarded to a queue in a timeslot. Is it the case in equations (7)-(9) as well? now it's (3)-(5)} \esr{ - Not exactly, the Transition probabilities we have works for any $a$ not just 0 or 1. Anyways, for eliminating the ambiguity, I have mentioned what values can $s,s',a,b$ take in Lemma 1. See Lemma 1. }

Arrivals at time slot $T_i$ precede departures spanning from $T_i$ to $T_{i+1}$.
Following the transition at time $T_{i}$, if $A_n(T_{i}) = 1$, an extra job is admitted to the $n$-th queue,
and can be processed between $T_{i}$ and $T_{i+1}$ with the other jobs already in the queue. 
Therefore,
%
%
%
\begin{equation} \label{eq:sys_dynamics}
\begin{split}
    S_n(T_{i+1}) &= S_n(T_{i})  + A_n(T_{i})- D_{n}(T_{i}). 
\end{split}
\end{equation}
Then, $D_n(T_{i}) \leq S_n(T_i)+ A_n(T_i)$ and $S_n(T_i)=K \Rightarrow A_n(T_i)=0$. The number of   rejected jobs across all queues is  given by 
\begin{equation}
    R(T_{i})=\alpha N - \sum_{n=1}^N A_n(T_i). \label{eq:rejections}
\end{equation}

We denote by $B_n(T_i)$ an indicator variable representing the  service rate of the $n$-th queue, with the  values 0 and 1 indicating  that  the server   operates at a low rate   and at a high rate, respectively.   It is assumed that the service rate  remains constant  between two arrivals. We let $q(b)$ denote the probability of a service completion, at a given slot, when the controller sets the service rate indicator to $b$. The service completion probabilities corresponding to the low and high service rates are given by
$q(0)= \underline{b}$ and  $q(1) = \overline{b}$.

The \textbf{transition probabilities} between states of a queue are defined by $P$.  To derive those probabilities, we let $s$ be  the current state,   $a$  be the current job allocation action, and   $s'$   be the next state.

Next, we consider the interval comprised of $\tau$ time slots, wherein the system switches from state $s$ to state $s'$.
We account for two possible scenarios, varying on how jobs are serviced across those slots: 

  \textbf{Concurrent job service per slot, per queue  (CJS): }  Each pending job at each queue can be  served  concurrently with other jobs. 
In this case,  more than one job can be served per queue per time slot. This corresponds to a setup wherein a cluster of servers, or a single server with multiple cores, is available for each queue.   

     \textbf{Single job service per slot, per queue  (SJS): } Only one job can be processed per queue at a given time slot. 
In particular, at each time slot, we assume that \textbf{head-of-line (HOL)} jobs across queues are candidates to be served.  This corresponds to a setup wherein a single core is available to serve each queue per time slot. 



\ReviewerC{Further description of Eq.(6) would enhance readability, as it appears somewhat abrupt. I understand this equation serves as foreshadowing for Lemma 1 but providing additional context or explanation regarding its relation to the transition probabilities and the Bernoulli distribution would be beneficial.} \esr{- Done. See the next paragraph.}

Let $\mathbb{P}(\mathcal{B}_{m,p}=\ell)$ denote the probability mass function of the binomial distribution, indicating the probability  of   $\ell$ successes in $m$ independent Bernoulli trials. Then, $\mathbb{P}(\mathcal{B}_{m,p}=\ell)=
    \binom{m}{\ell}  p^\ell (1-p)^{m-\ell}$
where $\binom{m}{\ell}=m!/((m-\ell)!\ell!)$ for $0 \leq \ell \leq m$, and 0 otherwise. We use the above notation to describe the     distribution of the number of jobs processed in a given time period. The 
transition probabilities for CJS and SJS  are presented in the following lemma: 
\begin{lemma} \label{lemma:probs}
    Let $s,s' \in \{0,1,...,K\}, a \in \{0,1\}$ and $b \in \{0,1\}$. Then, the transition probabilities $\mathbb{P}(S_n(T_{i+1}) = s' \mid S_n(T_{i}) = s, A_n(T_{i}) = a, B_n(T_{i}) = b) 
    $, denoted by $P_{s,s',a,b}$  are given by:
      \begin{align} \label{eq:pssab}
   P_{s,s',a,b} &=  \sum_{\tau=1}^{\infty} (1-p)^{\tau-1} p P_{s,s',a,b,\tau} 
   \end{align} 
   where
   \ifthenelse{\boolean{compactcameraready}}
   {
   $P_{s,s',a,b,\tau} := \mathbb{P}(S_n(T_{i+1}) = s' \mid S_n(T_{i}) = s, A_n(T_{i}) = a, B_n(T_{i}) = b, T_{i+1} - T_{i}=\tau).$
   }{
       \begin{equation}
           \label{eq:pssabtau}
           \begin{split}
               &P_{s,s',a,b,\tau} := \mathbb{P}(S_n(T_{i+1}) = s' \mid S_n(T_{i}) = s, A_n(T_{i}) = a,\\
               & \quad B_n(T_{i}) = b, T_{i+1} - T_{i}=\tau).\\
            \end{split}
    \end{equation}
}
    Under CJS,
    \begin{equation}
        P_{s,s',a,b,\tau} = \mathbb{P}\left(\mathcal{B}_{s+a,1-(1-q(b))^\tau}=s+a-s'\right).
    \end{equation}
    Under SJS,
    \begin{equation}
    P_{s,s',a,b,\tau} =    \left\{
    \begin{aligned}
        &  
        \mathbb{P}\left(\mathcal{B}_{\tau,q(b)} \ge s+a\right),   && \text{if } s'=0   \\ 
         &  
        \mathbb{P}\left(\mathcal{B}_{\tau,q(b)} =s+a-s'\right),   && \text{if } s' > 0.   
        %
    \end{aligned}
    \right.  
    \end{equation}

\end{lemma}

\ifthenelse{\boolean{compactcameraready}}{


}{

\begin{proof}

\ReviewerC{Regarding the proof of Lemma 1, it would be helpful to clarify whether the event ${\delta T_i = \tau}$ and the event ${S_n(T_i) = s, A_n(T_i) =a, B_n(T_i)=b}$ are considered independent.}

Let $s,a,b$ denote the state of the queue, job allocation action and service rate action at time $T_i$. Let $s'$ denote the state of the queue at time $T_{i+1}$. We denote  $\mathbb{P} (S_n(T_{i+1}) = s' \mid S_n(T_{i}) = s, A_n(T_{i}) = a, B_n(T_{i}) = b)$ with some abuse of notation as $P(s'|s,a,b)$. Let $\Delta T_i = T_{i+1}-T_i$.
\ReviewerC{The second formula after Equation (9) contains an extra equal sign.}
Then,
\begin{equation*}
    \begin{split}
        \mathbb{P} ( s' \mid s, a, b) 
        &= \sum_{\tau=1}^\infty \mathbb{P} (s', \Delta T_i  = \tau \mid s, a, b)    =   \\
        & = \sum_{\tau=1}^\infty \mathbb{P} (\Delta T_i  = \tau) P_{s,s',a,b,\tau}
    \end{split}
\end{equation*}
where the conditional probability $P_{s,s',a,b,\tau}$ is defined in~\eqref{eq:pssabtau}. 



\emph{SJS. } We now derive $P_{s,s',a,b,\tau}$ for SJS. 
Across $\tau$ time slots,  we aim at characterizing the  occurrence of $s+a-s'$ successes, if $s'>0$, and at least  $s+a$ successes, if $s'=0$, with the success probability at each time slot being given by $q(b)$.  
Indeed, given an interarrival time between jobs of $\tau$, the number of jobs  processed in this time period, at the considered queue,  is $s+a-s'$, where $s+a-s'  \leq \tau$.  The service rate $b$ entails a probability of successful job completion per slot given by  $q(b)$. Hence, if $s'>0$, the probability of processing $s+a-s'$ jobs during  $\tau$ time slots is given by the binomial distribution $\mathbb{P}\left(\mathcal{B}_{\tau,q(b)} = s+a-s'\right)$. 
Note that this probability is strictly positive if $0 \leq s+a-s' \leq \tau$, and equals 0 otherwise. The case $s'=0$ is similar, noting that even if more than $s+a$ potential successes occur, only $s+a$ jobs can be served.


\emph{CJS. }  We now derive the $P_{s,s',a,b,\tau}$ for CJS.  Recall that under CJS  multiple jobs can be processed per queue at a given time slot.
With  $s+a-s'$ jobs served during an interval of $\tau$ time slots, a binomial distribution with  $s+a-s'$ successes among $s+a$ trials  is used to characterize the probability of transitioning from $s$ to $s'$, where  each potential  service is completed   with  probability  $q(B_n(t))$ per slot.
Indeed, at each time slot, each of the pending jobs is processed with  probability $q(b)$. Therefore, the probability that a job is  served during a period of $\tau$ time slots is $1-(1-q(b))^\tau$. The probability that $s+a-s'$ jobs are served is $\mathbb{P}\left(\mathcal{B}_{s+a,1-(1-q(b))^\tau}=s+a-s'\right)$. Note that this probability is strictly positive if $s' \leq s+a$, and equals 0 otherwise. 
\end{proof}

}


\label{subsec:transition-p}

In Lemma \ref{lemma:probs}   we are considering the system dynamics from the perspective of one queue. 
 In particular, 
 the dimensions of the transition matrix  $P$ are given by  the maximum values of  $(s, s', a, b)$, which are $ (K+1, K+1, 2, 2)$.  The combinatorial growth in the state space cardinality occurs when examining a system comprised of multiple queues, as detailed next.








\section{Problem formulation and solution}
\label{sec:problem-formulation}

\ReviewerB{The paper is not well written, important steps should be explained better, both in the proposed solution and in the numerical results.}
\esr{How to address this issue? - Alex told we can ignore}

In this section, we begin by introducing our cost functions and constraints (Section~\ref{subsec:opt-problem}) and then  formulate our problem as a weakly coupled MDP (Section~\ref{subsec:lp-formulation}). The solution of a weakly coupled MDP is   known to be an NP-hard problem  when constraints are set on sample paths~\cite{gast2021lp}. Therefore,  we relax  the problem to consider constraints on expectations, amenable to be solved with a manageable LP  (Section~\ref{subsec:relaxed-lp}). The latter, in turn,  is  asymptotically optimal, in the sense that as  the number of queues grows to infinity its solution converges to that of the original problem~\cite{lp-policy}. We proceed to expand the LP problem into its Lagrangian form and introduce a two-timescale stochastic approximation algorithm to tackle the problem in cases where transition probabilities are unknown, and the optimal policy must be learned in an online fashion (Section~\ref{subsec:2ts-sa}). 

\subsection{Optimization problem}
\label{subsec:opt-problem}

We introduce the cost functions and constraints to formulate the optimization problem.

The \textbf{cost functions} comprise two components:

\begin{enumerate}
    \item The \textbf{storage and processing costs}, comprising the \emph{storage cost} $C_s(S_n(t))$ which, due to Little law,  is  proportional to delays, and the \emph{processing cost} $C_p(B_n(t))$, which is proportional energy consumption. Both costs are assumed to be convex and increasing;
    \item The \textbf{job rejection cost}, given by $\gamma R(t)$ (see~\eqref{eq:rejections}), where  $\gamma$ is a positive factor capturing the weight of job rejections when compared against other costs. Note that as $\alpha N$ is a constant,   minimizing job rejection costs  $\gamma R(t)$ is equivalent to minimizing  $-\gamma\sum_{k}A_n(t)$.
\end{enumerate}

The \textbf{constraints} also comprise two components:

\begin{enumerate}
    \item For the \textbf{storage cost}, the allocation of jobs to queues must not exceed the batch size of $\alpha N$ jobs, where $\alpha \in (0,1)$;
    \item For the \textbf{energy cost}, the total number of queues operating at a high processing rate $\overline{b}$ must not exceed $\beta N$, where $\beta \in (0,1)$.
\end{enumerate}

Collectively, these elements define the optimization problem, as detailed in the subsequent subsection.

\subsection{Problem formulation}
\label{subsec:lp-formulation}

Before we state the optimization problem, we introduce some key notations.
Let   $Y\mysup_{s,a,b}(t)$ be the fraction of queues in state $s$ and subject to actions $a,b$ at time $t$.
\ReviewerB{Please define better what a policy is. Now it feels like the actions are defined not only by the state of the queues and the arrivals but also by the time index. TRUE IS IT EXPLICIT IN OUR PAPER.} \esr{Done. See the first line of optimization problem subsection below.}
We consider a finite time horizon $T$, so that $0 \leq t \leq T-1$. 
Let $Y$ be a vector of random variables,  $Y\mysup=(Y\mysup(1), \ldots, Y\mysup(t), \ldots,  Y\mysup(T))$, where each $Y\mysup(t)$   is   a random vector comprised of elements $Y\mysup_{s,a,b}(t)$.

 Let $M\mysup(t)=(M\mysup_s(t))_{s \in \mathcal{S}}$ be a vector whose  $s$-th entry corresponds to the fraction of queues containing $s$ jobs at time $t$.  Note that  $M\mysup_s(t)=\sum_{a,b} Y\mysup_{s,a,b}(t)$.
We denote by $m_s^0$ the initial system state,   $m_s^0 = M_s\mysup(0)$. 
It corresponds to   the fraction of queues in state $s$ at time $t=0$ for all $s\in\cS$.   Unless  otherwise noted, we assume  that   queues begin in an empty state, i.e., $m_0^0=1$.

\subsubsection{Optimization problem}

Let $\Pi$ be a decision rule  (or policy) that depends on both time and the state of the system. Then,  
$\Pi_t: m \mapsto y$ and 
 $\Pi_t(M\mysup(t))  \mapsto Y\mysup(t) $. The variables $Y\mysup(t)$ are then set as $Y\mysup(t) = \Pi_t(M\mysup(t))$. 
The optimization problem can now be formulated as follows: \vspace{0.03in}

\noindent\textsc{Problem with inequality constraints on sample paths:}
\begin{mini!}|s|{\pi}{ \E
\nonumber    \sum_{t=0}^{T-1}  \left( \sum_{s,a,b}(C_s(s) {+} C_p(b))Y_{s,a,b}\mysup(t) +\gamma\sum_{s,b}Y_{s,0,b}\mysup(t) \right)}{\label{eq:LP-WCMDP}}{}
    \addConstraint{\text{Queues evolve according to a Markov process} \nonumber }
    \addConstraint{
    \mathbb{P}(\bm{s}(t{+}1){|} \bm{s}(t), \bm{a}(t),  \bm{b}(t)){=} {\prod_{n{=}1}^N} {P}_{{s_n}(t),{s_n}(t{+}1),{a_n}(t), b_n(t)} 
    \label{eq:toimprove}}
    \addConstraint{\sum_{s,b}Y\mysup_{s,1,b}(t)\leq \alpha,  \quad  \sum_{s,a}Y\mysup_{s,a,1}(t)\leq \beta \label{eq:constrbeta}}
     \addConstraint{\sum_{a,b}Y\mysup_{s,a,b}(0) = m_s^0, \quad   Y\mysup_{K,1,b} = 0 \label{eq:constraintini}}
    \addConstraint{Y\mysup_{s,a,b}(t) \geq 0,  \forall t\in \{ 0, \ldots, T-1 \},\forall s\in \{0, \ldots, K\}\nonumber}
    \addConstraint{\quad \forall a \in \{0,1\}, \forall b \in \{0,1\}}   \label{eq:nonneg}
\end{mini!}
The objective function captures   costs introduced in the previous section.  State dynamics in~\eqref{eq:toimprove} follow the transitions derived in Lemma~\ref{lemma:probs}. Inequality constraints~\eqref{eq:constrbeta}  capture constraints on the workload and energy. Boundary conditions, i.e., initial conditions and droppings due to full buffers,  are captured by~\eqref{eq:constraintini},  and non-negativity constraints by~\eqref{eq:nonneg}. It is possible to write the above problem as an LP. However, the number of optimization variables grow combinatorially with buffer size and number of queues. 
\ifthenelse{\boolean{compactcameraready}}{
    Please refer to \cite{our_github_version} for more details. In what follows, we resort to a relaxation of the constraints, which allows us to formulate the relaxed problem as a manageable LP.
}{



\esr{The above was discussed in the call. However, I think the complexity is well explained by Daniel in the below text, which we have commented off. Shall we put the version we submitted to IFIP in arxiv and refer to that version for the details on complexity instead of referring to [23] The Lp-update policy paper?}

\subsubsection{Complexity}
\label{subsubsec:lp_complexity}
It is possible to write the above problem as a very large LP.   
Let set $\mathcal{Y}(m)$ contain all vectors $\realizationY$ that satisfy~\eqref{eq:constrbeta},  and such that the fraction of queues in state $s$ is given by $m_s$, with $m=(m_0, m_1, \ldots, m_{K})$.

The variables in our LP are probabilities $p_{t,\realizationY}$ for all admissible $\realizationY$, $\realizationY \in \mathcal{Y}(m)$,  where   $p_{t,\realizationY} = \mathbb{P}( Y(t) = \realizationY)$. We also let  $q_{t,m}=\mathbb{P}(M(t)=m)$, and $P(m | \realizationY)=\mathbb{P}(M(t+1)=m | Y(t)=\realizationY)$.   
  Then,   equation~\eqref{eq:toimprove} is captured in the LP  through the following constraint   
$
q_{t+1,m} = \sum_{\realizationY} p_{t,\realizationY}  P( m | \realizationY ).$ 
The Markovian evolution is given by   $P(m | \realizationY)$, which denotes the probability of the system transitioning from state-action pair $\realizationY$ to state $m$.  
Inequalities~\eqref{eq:constrbeta}  are captured in the LP by
$ q_{t,m} =   \sum_{\realizationY \in \mathcal{Y}(m) } p_{t,\realizationY}$,  indicating that $\realizationY$ must be feasible for $p_{t,\realizationY}$ to contribute to $q_{t,m}$, and relating the probabilities of state-action vectors $\realizationY \in \mathcal{Y}(m)$ to the probability of state vector $m$.

The issue with the above formulation arises as the number of possible values for $\realizationY$ quickly becomes very large.  
In fact, the primary challenge in solving the optimization problem above lies in its state space cardinality.  Note that the random vector $Y(t)$ can be instantiated in up to $\binom{4(K+1)-1+N}{N}$ possible ways, each corresponding to a decision variable $p_{t,\hat y}$ to be determined. The binomial term corresponds to the number of ways of dividing the $N$ queues into  $4(K+1)$ buckets, each bucket corresponding to one of the possible instantiations of $(s,a,b)$.   
%
Despite the large number of variables  $p_{t,\hat y}$ to be determined, the variables are all continuous, in the range between 0 and 1. Hence, the problem is in the realm of LPs.
 

As the constraints~\eqref{eq:constrbeta} couple the states of all queues, they preclude the possibility of dividing the problem into $N$ problems of linear size each, rendering its solution unwieldy. In what follows,   we resort to a  relaxation of the constraints, which allows us to formulate the relaxed problem as a manageable LP.

\ReviewerD{In Section IV.B, it would be helpful to give a numeric illustration at the end of the section to show how the problem size grows. This could be also shown in a table for different values of the key parameters. This would help appreciate the following section on the Relaxed problem, along with a numeric illustration. The tables in Sections IV.B and IV.C could also be combined into a single table as an alternate way to present this information.}

}

\subsection{Relaxed problem becomes a manageable linear program}
\label{subsec:relaxed-lp}

Next, we consider a relaxed version of the  problem introduced in the previous section.  As discussed above,  the key difficulty in solving the posed problem arises  from   constraints~\eqref{eq:constrbeta}
which couple all the queues together. To overcome the challenge, we relax the constraints. The common approach is to relax the constraints by posing a problem where they only need to be met in expectation. Let 
\begin{equation}
    y_{s,a,b}(t)=\E[Y\mysup_{s,a,b}(t)].
\end{equation}
Let   $\pi_H(t)$ and $\pi_A(t)$ be the probability of using the high service rate  and the probability of accepting a job, respectively,
\begin{equation}
    \pi_H(t) = \sum_{s=0}^{K} \sum_{a=0}^1 y_{s,a,1}(t), \quad   \pi_A(t) = \sum_{s=0}^{K-1} \sum_{b=0}^1 y_{s,1,b}(t).  \label{eq:pia} 
\end{equation}
%
The relaxed problem  to derive the optimal policy is an LP: \vspace{0.03in}


\noindent\textsc{Problem with inequality constraints on expectations:}
\begin{mini!}|s|{y}{
   \nonumber \sum_{t=0}^{T-1} \left( \sum_{s,a,b}(C_s(s)+C_p(b))y_{s,a,b}(t) + \gamma (1-\pi_A(t))  \right)
}{\label{eq:LP-bandit}}{}
    \addConstraint{
        \sum_{a,b}y_{s,a,b}(t+1) = \sum_{s',a,b}y_{s',a,b}(t)P_{s',s,a,b}}
    \addConstraint{
        \pi_A(t) \leq \alpha, \quad \pi_H(t) \leq \beta \label{eq:alphabetaexpcon}}
    \addConstraint{
        \sum_{a,b}y_{s,a,b}(0) {=} m_s^{0}, y_{K,1,b}(t) {=} 0, y_{s,a,b}(t) \geq 0}
    \addConstraint{\forall t\in \{0,\ldots,T-1\},\forall s\in \{0,\ldots, K\} \nonumber }
    \addConstraint{ \nonumber  \forall a \in \{0,1\}, \forall b \in \{0,1\}. \nonumber}
\end{mini!}


The solution to the relaxed LP converges asymptotically to the solution of the original LP problem as  $N$ grows to infinity~\cite{lp-policy}. However, applying the solution of the relaxed LP directly to the original system may not be feasible under the current system configuration.  To  address this limitation, in \cite{lp-policy} the authors suggest to solve the LP taking  the current system configuration as the initial state, and the  remaining time of interest as the horizon $T$ \cite[Algorithm 1]{lp-policy}.   An alternative approach is proposed in~\cite{yan2023certainty}. 
In essence, those works provide ways to map a solution to the LP with  constraints on expectations~\eqref{eq:alphabetaexpcon} to a solution to the stricter LP with   constraints on samples paths~\eqref{eq:constrbeta}.  Given that such mappings are known to exist, in the remainder of this paper we focus on the  problem with constraints on expectations, noting that constraints on expectations are of interest by itself~\cite{altman2021constrained}, and leaving a detailed analysis of the mapping between constraints on expectations and   sample paths as subject for future work.


Handling the above LP    becomes 
challenging when   transition probabilities $P_{s',s}^{a,b}$ 
   are unknown. One of our contributions lies in solving the relaxed LP in the presence of unknown transition probabilities. To this aim, we derive the Lagrangian of the relaxed LP and employ a two-timescale stochastic approximation scheme to tackle the LP in an online fashion. Our approach is further elaborated in the following section, with numerical experimental results presented in Section~\ref{sec:experiments}.




\subsection{A two-timescale stochastic approximation algorithm}
\label{subsec:2ts-sa}

Next, we  derive the Lagrangian for the relaxed LP formulation. Subsequently, we leverage the Lagrangian to propose a  two-timescale stochastic approximation algorithm to solve the problem introduced in the previous section. We denote the Lagrangian of the LP in \eqref{eq:LP-bandit} as $L(y,\lambda,\mu)$.  Here, {\color{black}$\lambda$ and $\mu$} are vectors of Lagrange multipliers corresponding to the {\color{black}inequality and equality    constraints in~\eqref{eq:LP-bandit}, respectively}. 
\begin{align}
    &L(y,\lambda,\mu;P, m^0) =\nonumber \\
    & \quad \sum_{t=0}^{T-1} \mathcal{C}(y, t) + \mathcal{D}_1(y,\mu, t; P,m^0)  +\mathcal{D}_2(y,\lambda, t), 
\end{align}
 $\mathcal{C}(y, t) = \sum_{s, a, b} (C_s(s) + C_q(b)) y_{s, a, b}(t) + \gamma \sum_{s, b} y_{s, 0, b}(t)$.

\setlength{\tabcolsep}{0pt}

\begin{figure*} \centering
     \begin{tabular}{ccc} \includegraphics[width=0.32\textwidth]{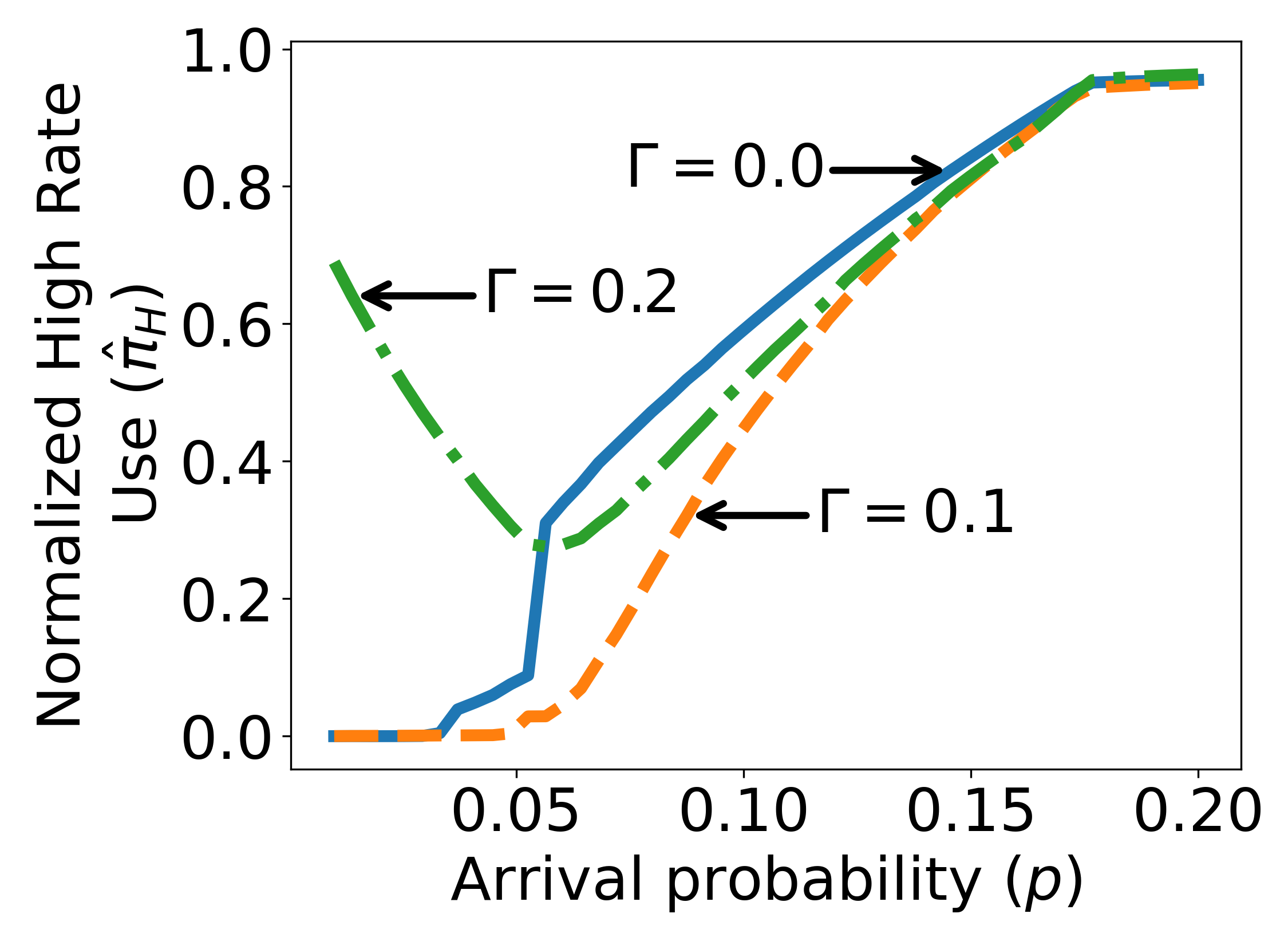} &  \includegraphics[width=0.32\textwidth]{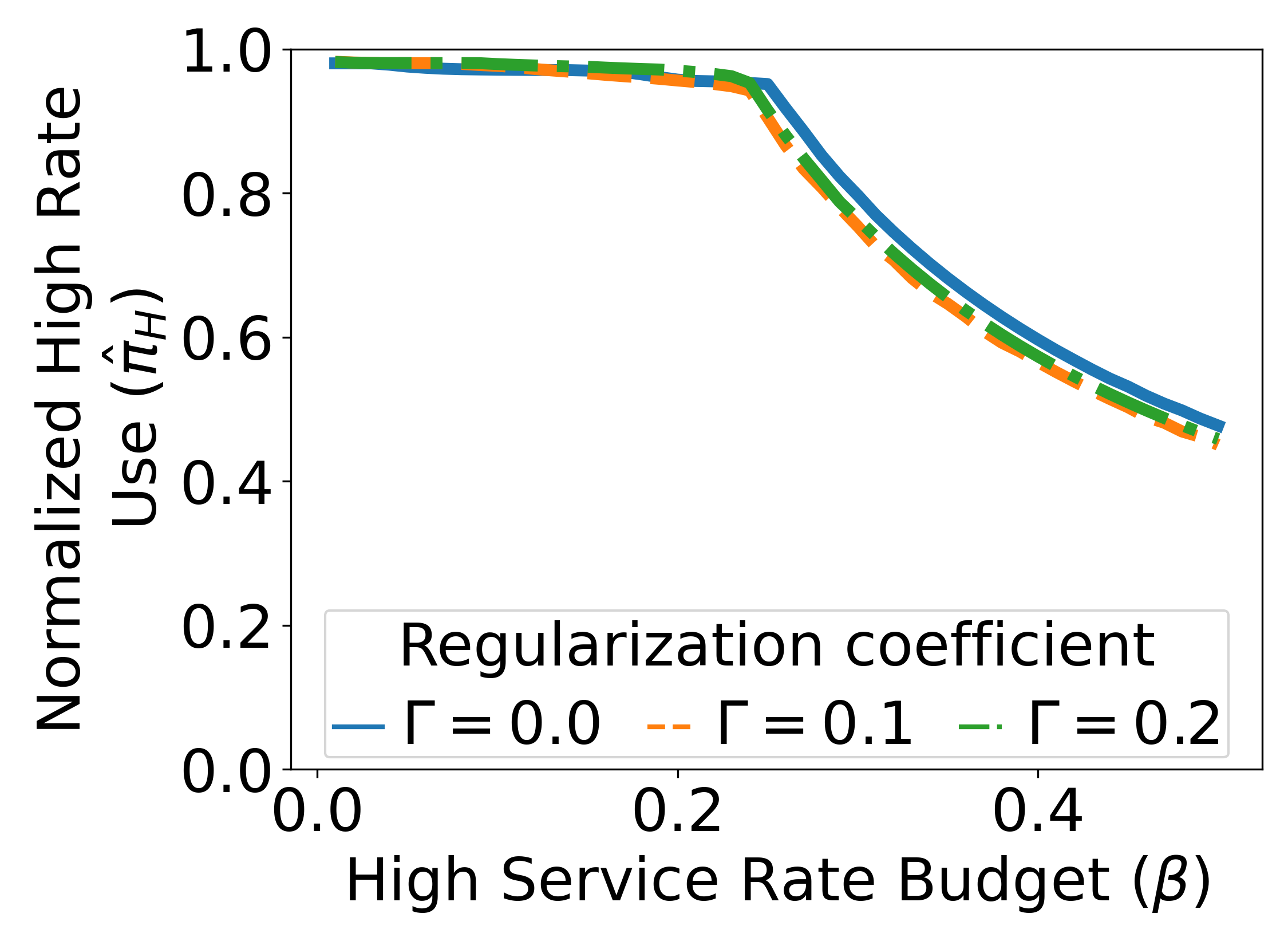}& \includegraphics[width=0.32\textwidth]{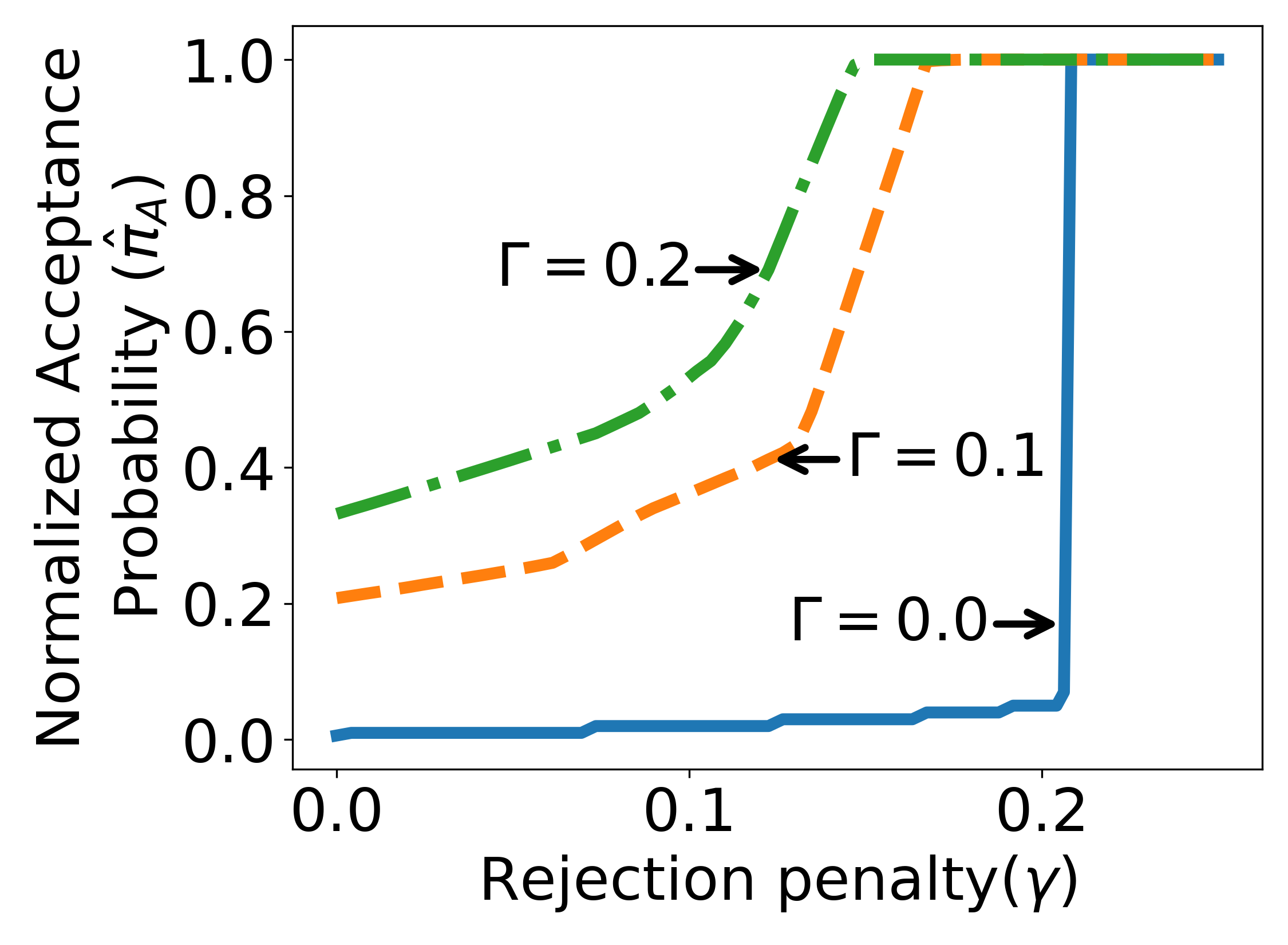}\\
     (a) & (b) &(c)
     \end{tabular}
        \caption{Analysis of the impact of (a) arrival probability $p$, (b) high service rate budget $\beta$, and (c)  job dropping cost $\gamma$ on the solution of LP-based policy~\eqref{eq:LP-bandit}. The figure also shows the impact of the regularization coefficient, $\Gamma$, on the optimal policy.}
        \label{fig:parameter_analysis}
        \vspace{-0.15in}
\end{figure*}

The Lagrangian terms corresponding to  equality and inequality constraints are denoted  by $\mathcal{D}_1(y,\mu, t; P,m^0)$ and $\mathcal{D}_2(y,\lambda, t)$, respectively. 
\ifthenelse{\boolean{compactcameraready}}{For their expressions, we refer the reader to  \cite{our_github_version}.
}{


\begin{align}
& \mathcal{D}_1(y, \mu,t;P,m^0) = \sum_s \mu_1^s \left( \sum_{a, b} y_{s, a, b}(0) - m^0_s \right)+  \nonumber \\  & \quad +\sum_b \mu^b_2(t) y_{K, 1, b}(t) +\nonumber  \\
&\quad+\sum_s \mu_3^s(t) \left( \sum_{a, b} y_{s, a, b}(t+1) - \sum_{s', a, b} y_{s', a, b}(t)P_{s', s}^{a, b} \right), \nonumber  
\end{align}
\begin{align}
    &  \mathcal{D}_2(y, \lambda, t) = \lambda_1(t)  \left( \sum_{s, b} y_{s, 1, b}(t) - \alpha \right) + \nonumber \\
    & \quad +\lambda_2(t) \left( \sum_{s, a} y_{s, a, 1}(t) -\beta \right)  - \sum_{s, a, b} \lambda_3^{s, a, b}(t) y_{s, a, b}(t).\nonumber
\end{align}
}
Since the Lagrangian exhibits linearity with respect to $y$, the derivative of the Lagrangian with respect to $y$ is constant. Therefore, we proceed by extending the previous formulation and introducing a regularization term to the problem. The regularized Lagrangian is expressed as follows:
\begin{equation} \label{eq:reg_lagrangian}
\tilde{L}(y,\lambda, \mu;  P, m^0) = L(y,\lambda, \mu; P, m^0) + \Gamma ||y||^2_2.
\end{equation}
Then, we minimize the Lagrangian with respect to $y$ and maximize with respect to the Lagrange multipliers to find a saddle point which corresponds to an optimal policy:  \vspace{0.03in}

\noindent\textsc{Problem with constraints as penalties:}

\begin{equation} \label{eq:regularizedproblem}
\max_{\lambda \geq 0, \mu} \min_y  \tilde{L}(y,\lambda, \mu; P, m^0).
\end{equation}
If all parameters are known, the above problem can be solved using a Primal-Dual Gradient Descent-Ascent (GDA) method~\cite{Gidel2020Variational,chenprimal,boyd2004convex}. However, in this work we assume that the transition probabilities $P$ is unknown. 
\ReviewerB{"In the absence of prior information about the transition matrix P, we resort to simulations to estimate P and learn the optimal policy" I do not understand this sentence. For the simulation the P matrix needs to be known, right? Also, when would this simulation be done? Off-line in the system design phase?}
To this aim, we perform stochastic gradient updates. While computing the Lagrangian, at each step,   instead of using the actual $P$ matrix, we use an estimate of $P$. We denote by $\hat{P}^{(k)}_{s,s',a,b}$ the estimate of $P_{s,s',a,b}$ (see~\eqref{eq:pssab}) at  iteration $k$, e.g., derived  from a simulator.  
The estimate $\hat{P}_{s,s',a,b}^{(k)}$ is obtained  by repeatedly performing actions $a, b$ in state $s$. Let $I$ be the number of collected observations. 
$\hat{P}^{(k)}_{s,s',a,b}$ corresponds to  the fraction of those observations that lead to   state $s'$.   Our final stochastic gradient iterates are:
\begin{eqnarray}
    y^{(k+1)} &=& y^{(k)} - \eta_1 \nabla_{y}\tilde{L}(y^{(k)},\lambda^{(k)},\mu^{(k)};\hat{P}^{(k)}, m^0)  \label{eq:y_update_P_hat} \\
    \lambda^{(k+1)} &=& \left[\lambda^{(k)} + \eta_2 \nabla_{\lambda}\tilde{L}(y^{(k)},\lambda^{(k)},\mu^{(k)}; \hat{P}^{(k)}, m^0)\right]_+ \label{eq:lambda_update_P_hat} \\
    \mu^{(k+1)} &=& \mu^{(k)} + \eta_2 \nabla_{\mu}\tilde{L} (y^{(k)},\lambda^{(k)},\mu^{(k)}; \hat{P}^{(k)}, m^0) \label{eq:mu_update_P_hat}
\end{eqnarray} 
The above description to solve the  stochastic problem corresponds to a  stochastic variant of GDA, referred to in the literature as Stochastic Gradient Descent-Ascent (SGDA). 
We let $y$ vary at a fast time scale, and $\lambda, \mu$ vary at a slow scale. We denote by $\eta_1$ and $\eta_2$ the learning rates of the fast and slow time scales, respectively. Note that, $[x]_+ = \max(0,x)$, enforcing  that $\lambda \ge 0$. Recall that the Lagrange multipliers $\lambda$ and $\mu$ correspond to   inequality and equality constraints, respectively. Therefore, after convergence of the above iterative procedure,   the fixed point satisfies the KKT conditions, including complementary slackness, as $\lambda \ge 0$, and corresponds  to  an optimal solution to the LP problem.

\section[Experiments]{Experiments}
\label{sec:experiments}

\ReviewerC{The experiments do not include real-world scenarios, which are crucial for validating the effectiveness of the proposed model and algorithms. Including real-world scenarios would enhance the practical relevance and applicability of the findings. Analyzing convergence characteristics can be challenging, so the lack of analysis of algorithm convergence characteristics is understandable. However, if the paper is accepted, I hope that the algorithm can be validated in real-world scenarios. ADD A LINE IN THE CONCLUSION}

\ReviewerD{The title refers to online learning, but the results section covers just convergence. There is no discussion on how quickly the problem is solved depending on the size of the problem to make any decision through online learning to do load balancing and auto-scaling. WE CAN TAKE A DECISION AT EVERY ITERATION OF THE ALGORITHM, AND ONE ITERATION IS CLOSED TO BE INSTANTENOUS. THEREFORE ONLINE.}
\esr{Where to add this sentence? In Introduction under online learning contribution or in the end of paragraph after eq 15? - Alex in introduction.}

In this section we investigate properties of   load balancing and auto scaling strategies, by analyzing  the impact of different parameters on the optimal policy (Section~\ref{sec:optpol}). We  also numerically investigate convergence properties of the considered algorithms (Section~\ref{sec:evalalgos}). 

For our numerical evaluation, the storage cost is given by
$C_s(s) = s/K$
and the processing costs are given by
$C_p(b) = 2 (1+q(b)).$ 
In particular,  it follows from Little's law that the linear storage cost is proportional to the  delay.

\subsection{How does the optimal policy behave?} \label{sec:optpol}

Next, we report properties of the optimal policy, indicating how different parameters impact actions.  To this goal, we consider the policy obtained through the solution of~\eqref{eq:LP-bandit}, i.e., the  
\textsc{Problem with inequality constraints on expectations}, leveraging CVXPY for that matter. 



\subsubsection{What are the  impacts of different parameters on the optimal policy?} We consider the impact of $p$, $\beta$ and $\gamma$ on the optimal policy. To this aim, 
recall that $\pi_A(t)$ and $\pi_H(t)$ refer, respectively, to the probability of accepting a job and the probability of choosing the high service rate at time $t$. 
We derive from~\eqref{eq:pia} the normalized quantities $\hat{\pi}_A$ and $\hat{\pi}_H$,  
$    \hat{\pi}_A = \frac{1}{\alpha T} \sum_{t=0}^{T-1} \pi_A(t), $ and $ 
     \hat{\pi}_H = \frac{1}{\beta T} \sum_{t=0}^{T-1} \pi_H(t). $ 
Recall also  that $\pi_A(t) \leq \alpha$ and $\pi_H(t) \leq \beta$.  Therefore, $0 \leq \hat{\pi}_A \leq 1 $ and $0 \leq \hat{\pi}_H \leq 1 $, with   $\hat{\pi}_A =1$ and ${\hat\pi}_H(t) =1$  if the  corresponding inequality constraints are active. 
%

We let  $\alpha=0.5$, $q(0)=0.05$, $q(1)=0.1$, $K=10$ and $T=100$, 
under the CJS scenario.
In our reference setup, we also let  $p=0.14$, $\beta=0.3$ and $\gamma=10$. The latter three parameters are varied according to the experimental goals.  Those values are selected to simplify   presentation, allowing us to illustrate   insights by varying a parameter, while maintaining all others fixed, as indicated in the sequel.  The discussion that follows considers $\Gamma=0$ (solid lines in Figure~\ref{fig:parameter_analysis}).  The impact of $\Gamma$ is pointed in the end of this  section, motivated by  SGDA. 

 \textbf{Arrival rate up, high rate use rises.} Figure~\ref{fig:parameter_analysis}(a) shows that as job arrival probability ($p$) increases, overall system utilization and the likelihood of using the high service rate increase. This implies that with more jobs, there's a greater chance of assigning a high service rate to minimize dropping  probability and delay costs. Once $p$ hits a threshold (e.g., $p \approx 0.2$ in this case), the high service rate constraint becomes active, preventing further increase  of $\hat{\pi}_H$ (recall that $\hat{\pi}_H$ is   normalized by the high service rate budget $\beta$).

\textbf{Budget up, high rate use steady.} Figure~\ref{fig:parameter_analysis}(b) illustrates the impact of the high service rate budget ($\beta$) on its allocated fraction ($\hat\pi_H$). When the constraint is stringent ($\beta \leq 0.28$), nearly the entire budget is utilized ($\hat\pi_H \approx 1$). Conversely, with a relaxed constraint, i.e., as $\beta$ increases,  the allocated fraction $\hat\pi_H$ 
decreases, and the high-rate use $\beta \hat{\pi}_H$ remains roughly  steady. This suggests that increasing the budget for high service rate doesn't necessarily lead to increased resource consumption due to associated energy costs affecting the objective function. The motivation behind limiting high service rate   is driven by   budget constraints ($\beta$) and the need to manage energy expenditure reflected in the objective function's cost term ($C_p(b)$). Increasing $\beta$ beyond a certain threshold leads to a decrease in normalized high rate use $\hat{\pi}_H$ due to $C_p(b)$.

 \textbf{Costlier drops, fewer rejections. } 
As the cost associated with dropping jobs increases,  the normalized probability of job acceptance shows a phase transition (see Figure~\ref{fig:parameter_analysis}(c)). Up to $\gamma=0.2$, the normalized acceptance probability remains roughly 0, meaning that almost all jobs are rejected.  Indeed, if the rejection cost is low, it is beneficial to keep the system almost always empty, to avoid incurring energy and storage costs. However, as $\gamma$ surpasses the threshold 0.2, there is a phase transition,  and the normalized acceptance probability remains stable at 1  for $\gamma > 0.2$.  For large enough dropping costs, the acceptance budget should be fully utilized.

\textbf{Regularization coefficient. }  Up until now, we considered $\Gamma=0$. In the next section, we consider $\Gamma >0$, as required by GDA and SGDA.  
Figure~\ref{fig:parameter_analysis} shows how, as $\Gamma$ increases, the behavior of the optimal policy diverges from that obtained with $\Gamma=0$.   For instance, we see a  smooth  transition of $\hat{\pi}_A$  as a function of $\gamma$, for  $\Gamma \in \{0.1,  0.2\}$.  
This illustrates that one must carefully set $\Gamma$ while relying on SGDA, noting that a  detailed sensitivity analysis  of $\Gamma$ is   subject for future work.  

\subsection{How do the proposed algorithms behave?}

\label{sec:evalalgos}

Next, we consider convergence properties.
Our reference setup  consists of $p = 1/2$, $\alpha = 1/2$, $\beta = 1/2$, $\gamma = 100$,  $(\underline{b} = 0.4,  \overline{b} = 0.8)$ and $T=10$. We let  $\Gamma = 0.5$. $K$ is varied between 1, 2, 4 and~8. 
Under SGDA, we collect  a mini-batch of $I=10$ observations before  each gradient update, 
 with stepsizes $\eta_1 = 0.1$ and $\eta_2 = 0.01$.

\label{subsec:projection}

Recall that we leverage the gradient descent ascent algorithm (GDA) and its stochastic version (SGDA) to find solutions to the load balancing and auto scaling problem, by finding the saddle point associated with its Lagrangian~\eqref{eq:reg_lagrangian}. While GDA uses full information, SGDA learns the transition probabilities and the optimal policy concurrently.
In the rest of this section,  we compare GDA, 
SGDA, given by~\eqref{eq:y_update_P_hat}-\eqref{eq:mu_update_P_hat} and the LP solution obtained with  CVXPY, given by~\eqref{eq:LP-bandit}.

Firstly, our analysis focuses on the convergence of  GDA towards the optimal solution of the LP problem in~\eqref{eq:LP-bandit}. 
The convergence reveals that the Frobenius norm of the  difference between the GDA output and the optimal solution diminishes, approaching  zero after
20,000 iterations, 
for buffer sizes ($K$) ranging from 1 to 8. 
It's noted that prior to stabilization, the solution discrepancy exhibits oscillatory behavior. This is typical of GDA, wherein disparities in learning rates used for minimization and  maximization cause some  updates to be too aggressive compared to   others,   leading  to   overshooting and thus inducing oscillatory behavior    towards convergence~\cite{Gidel2020Variational}. 

Both CJS and SJS show similar trends, with SJS oscillations  reaching slightly larger values. This occurs because SJS costs can vary across a broader range of values than CJS, as SJS relies on a single service per slot whereas CJS can  concurrently serve multiple jobs  at a given slot.    Additionally, an increase in buffer size ($K$)  decreases  GDA's convergence speed, due  to the larger cardinality of the set of control variables.


Secondly, we contrast the  convergence behaviors of GDA and  SGDA.  
In particular, we focus on the Frobenius norm of the difference between the solutions found by the two algorithms at each simulation epoch. 
Drawing from the theory of stochastic approximation   (see Chapter 9 of~\cite{borkar2009stochastic}), SGDA's trajectories are expected to  mirror that of GDA.   Our results indicate  a close agreement between SGDA and GDA  after
35,000 iterations.
A formal treatment of the convergence  of  SGDA, leveraging~\cite{borkar2009stochastic}, is left as  subject for future work.

\section{Conclusion} 
\label{sec:conclusion}

We approached the challenge of load balancing and auto scaling across parallel queues through the lens of a weakly coupled MDP. Drawing on recent advancements in solving such systems, we proposed an LP-based policy  and developed a novel online learning algorithm to refine this policy when parameters are unknown and may change dynamically. Our numerical experiments shed light on   policy behaviors, including a phase transition in job acceptance probability as the dropping cost grows. The experiments also allow  us to assess  the efficacy of our online learning algorithm. 
Future work includes a real-world reality-check of the quality of the policies obtained through the proposed LP. 
To ensure reproducibility we make code and experiments publicly available~\cite{our_github_version}.


\textbf{Acknowledgements: }
 Eshwar was supported by the Prime Minister’s Research Fellowship (PMRF).  
G. Thoppe's research was supported in part by DST-SERB's Core Research Grant  CRG/2021/008330,  Walmart Center for Tech Excellence, and   the Pratiksha Trust Young Investigator Award.  This work was partially supported by CAPES STIC AMSUD project RAMONaaS,  CNPq and FAPERJ through grants 315110/2020-1, E-26/211.144/2019 and E-26/201.376/2021.




    
    


\bibliographystyle{unsrt} 

\bibliography{Main-new-6pages} 





\end{document}